\overfullrule=0pt
\font\twelvebf=cmbx12
\font\ninerm=cmr9
\nopagenumbers
\overfullrule=0pt
\baselineskip=18pt
\line{\hfil CCNY-HEP 96/12}
\line{\hfil RU-96-13-B}
\line{\hfil September 1996}
\vskip .3in
\centerline{\twelvebf Gauge invariance and mass gap in (2+1)-dimensional}
\centerline{\twelvebf Yang-Mills theory}
\vskip .4in
\baselineskip=14pt
\centerline{\ninerm DIMITRA KARABALI}
\vskip .05in
\centerline{Physics Department}
\centerline{Rockefeller University}
\centerline{New York, New York 10021}
\vskip .4in
\centerline{\ninerm V.P. NAIR}
\vskip .05in
\centerline{Physics Department}
\centerline{City College of the City University of New York}
\centerline{New York, New York 10031}
\footnote{}{E-mail addresses: karabali@theory.rockefeller.edu, 
vpn@ajanta.sci.ccny.cuny.edu }
\footnote{}{Invited talk at the Workshop on Low Dimensional Field Theories,
Telluride, Colorado, August 1996 (to appear in the proceedings).}
\vskip .5in
\baselineskip=16pt
\centerline{\bf Abstract}
\vskip .1in
In terms of a gauge-invariant matrix parametrization of the fields, we give an
analysis of how the mass gap could arise in non-Abelian gauge
theories in two spatial dimensions.
\vfill\eject

\footline={\hss\tenrm\folio\hss}
\def\bp {\bar p}
\def\dag {\dagger}

\def\bdel{\bar{\partial}}

\def\D {\Delta}
\def\bz {\bar{z}}
\def\12 {{\textstyle {1 \over 2}}}
\def\bG {\bar{G}}
\def\A {{\cal A}}
\def\C {{\cal C}}
\def\H {{\cal H}}
\def\O {{\cal O}}
\def\G {{\cal G}}
\def\vf {\varphi}
\def\S {{\cal S}}
\def\ra {\rangle}
\def\la {\langle}
\def\Tr {{\rm Tr}}
\def\E {{\cal E}}

\def\bD {{\bar D}}
\def\bA {{\bar A}}
\noindent{\bf 1. Introduction}

In this talk, we shall be discussing pure non-Abelian gauge theories,
i.e., with no matter fields, in two spatial dimensions. Specifically, 
we shall focus on a gauge-invariant analysis and the question of how a
mass gap could arise in these theories $^{1,2}$. As is well-known, this issue
was addressed long ago by Polyakov who considered an $SU(2)$-gauge theory
spontaneously broken down to $U(1)~^{3}$. If this were a $(3+1)$-dimensional
theory, there will clearly be monopole solutions. For the $(2+1)$-dimensional
theory, these ``monopole"
solutions can be considered as tunnelling configurations (or 
instantons) and a semiclassical analysis can be done by expanding the
functional integral around these configurations. This leads to confinement
and eventually a mass gap. It is believed that the results hold for the 
unbroken theory as well, although we go out of the regime of validity of 
the semiclassical expansion as the parameters are relaxed towards the 
unbroken phase. Here we would like to 
understand the mass gap directly in the unbroken theory. We will focus more on the geometry of the configuration 
space of the theory. After all, nonperturbative aspects of gauge theories 
are not well understood and exploring different points of view can be 
quite useful.

A question which might immediately arise is, why
$(2+1)$ dimensions ? Why not directly
analyze the more physical case of $(3+1)$ 
dimensions ? Apart from the fact that this is a conference devoted to
low dimensional field theories, there is a good technical reason why we hope to 
make more progress in $(2+1)$ dimensions, at least as a start. In
the latter case, in Hamiltonian analysis, many of the quantities of 
interest are defined in terms of two-dimensional fields and one can use 
known results from two dimensions, especially from conformal field theory.
To set the stage and define a framework for the discussion, let us consider
an $SU(N)$-gauge theory in the $A_0 =0$ gauge. The gauge potential
can be written as $A_i = -i t^a A_i ^a$, $i=1,2$, where $t^a$ are hermitian 
$N \times N$-matrices which form a basis of the Lie algebra of $SU(N)$ with
$[t^a, t^b ] = i f^{abc} t^c,~~{\rm {Tr}} (t^at^b) = {1 \over 2} \delta ^{ab}$. 
The Hamiltonian in this case is
given by
$$\eqalign{
\H = T~+V,~\ \ \ \
T={e^2\over 2} \int d^2 x ~ E_i^a E_i^a,~\ \ \ \
V= {1\over 2e^2} \int d^2x~ B^a B^a \cr 
E_i^a= \partial_o A_i^a ~= -i {\delta \over \delta A_i^a}, ~\ \ \ \
B^a= \12 \epsilon_{ij} (\partial_i A_j^a -
\partial_j A_i^a +f^{abc}A_i^b A_j^c)\cr}\eqno(1)
$$
The expectation value of the Hamiltonian for a physical state 
with wave function $\Psi[A]$ is given by
$$
\la \H \ra = \int d\mu ({\C}) ~ \left[ {e^2\over 2}
\left\vert{\delta \Psi \over \delta A}\right\vert^2 +{1\over 2e^2} B^2
 \vert \Psi \vert^2\right]\eqno(2)
$$

Wave functions for physical states are gauge-invariant and the integration 
in Eq.(2) is over all gauge-invariant field configurations. More precisely, 
let
$$
\eqalign{
{\A}&= \left\{ {\rm set ~of ~all ~gauge ~potentials} ~A_i^a(x)~ {\rm in} 
~{\bf R}^2 \right\}\cr
{\G}_* &= \left\{ {\rm set ~of ~all} ~g(x): {\bf R}^2 \rightarrow SU(N),
~~g\rightarrow 1 ~{\rm as} ~\vert \vec x \vert \rightarrow \infty \right\}\cr}
$$
${\G}_*$ acts on ${\A}$ in the usual manner of gauge transformations,
viz.,
$$
A_i^{(g)} = g^{-1} A_i g ~+ g^{-1} \partial_i g \eqno(3)
$$
The gauge-invariant configuration space $\C$ is then given by 
${\A}/{\G}_*$, viz., all gauge potentials modulo gauge transformations.

Many years ago Feynman tried to develop qualitative arguments regarding 
the wave functions and mass gap for this theory, 
similar in spirit to what he did so successfully 
for liquid Helium $^4$. It is easy to see that the ground state wave function can be
taken to be a real, positive function on $\C$ since the Hamiltonian operator
is real (and not just hermitian). The argument is similar to the standard 
argument of `no-nodes' for the ground state wave function of a quantum 
mechanical system without velocity-dependent potentials. 
It then follows that an
 excited state wave function $\Psi_{exc}$, being orthogonal
to the ground state, must be positive in some regions of $\C$ and negative in some
other regions. The kinetic energy term in Eq.(2), 
which is a gradient energy on $\C$, will roughly go like $1/L^2$ where $L$ is
some measure of the distance between the region where $\Psi_{exc}$ is positive
and where it is negative. Feynman suggested that, for the gauge theory,
 $L$ cannot be
made arbitrarily large and the gradient energy would have a definite nonzero
lower bound; this would be the mass gap. The reasoning is very similar to what
leads to a discrete spectrum for the Laplacian on a finite dimensional 
compact manifold. Indeed the kinetic energy term in Eq.(1) is the Laplacian 
on $\C$, but Feynman's argument cannot immediately be proved since
$\C$ is infinite dimensional and developing notions of 
compactness requires great care. Singer and others have
 given a careful formulation and 
discussion of some of the geometrical issues involved $^5$.

\vskip .1in
\noindent{\bf 2. A matrix parametrization of fields}

Fifteen years have passed since Feynman's paper and we have learnt many 
things about two-dimensional field theories and what we would like to do 
here is to reanalyze this line of reasoning. For this we need to understand the 
geometry of $\C= {\ A}/{\G}_*$ better; we shall need gauge-invariant 
variables for describing $\C$, we shall need to calculate the metric, 
volume element and Laplacian on $\C$. A good parametrization of the fields 
which allows the explicit calculation of $d\mu (\C )$ is a first step. We shall 
combine the spatial coordinates $x_1 ,x_2$ into complex combinations
$z=x_1 -ix_2,~{\bar z} =x_1+ix_2$; correspondingly we have
$A\equiv A_{z} = {1 \over 2} (A_1 +i A_2), ~~ 
{\bar A}\equiv A_{\bar{z}} =
{1 \over 2} (A_1 -i A_2) = - (A_z)^{\dagger}$.
The parametrization we use is given by
$$
A_z = -\partial_{z} M M^{-1},~~~~~~~~~~~~~~~~ A_{\bar{z}} = M^{\dagger -1} \partial_
{\bar{z}} M^{\dagger}
\eqno(4) 
$$
Here $M,~M^\dagger$ are complex $SL(N,{\bf {C}})$-matrices 
(for gauge group $SU(N)$). Such a parametrization is possible and is standard 
in many discussions of two-dimensional gauge fields. Indeed for any $A,~\bA$, 
it is easily checked that a choice of $M,~M^\dagger$ is given by
$$\eqalign{
M(x) &= 1 - \int G(x,z_1 ) A(z_1) +\int G(x,z_1)A(z_1) G(z_1,z_2)
 A(z_2) - ...\cr
&=~ 1 -\int_y  D^{-1}(x,y) A (y)\cr
M^\dag (x) &=
~1 - \int_y \bA  (y) \bD ^{-1} (y,x)\cr
}\eqno(5)
$$
(Here $D=\partial +A,~\bD =\bdel +\bA,~ 
\bar{G} (x,x') = {1 \over {\pi (z-z')}}, ~
G(x,x') = {1 \over {\pi (\bar{z} - \bar{z}')}},~{ \bar \partial}_x \bar{G} (x,y) = \partial _x G(x,y) ~= 
\delta ^{(2)} (x-y) $. There may be many choices for $M,~M^\dagger$; we shall discuss this question 
later.) From the definition (4), it is clear that a gauge transformation (3) 
is expressed in terms of $M, ~M^\dagger$ as
$$
M\rightarrow M^{(g)}=gM,~~~~~~~~~~~M^{\dagger (g)}=M^\dagger g^{-1} \eqno(6)
$$
for $g(x) \in SU(N)$. In particular, if we split $M$ into a unitary part $U$ 
and a hermitian part $\rho$ as $M=U\rho$, then $U$ is the `gauge part', 
so to speak; it can be removed by a gauge transformation and $\rho$ 
represents the gauge-invariant degrees of freedom. Alternatively, 
we can use $H=M^\dagger M =\rho^2$ as the gauge-invariant field 
parametrizing $\C$. Since $M\in SL(N,{\bf {C}}),~\rho$, and hence $H$, 
belong to $SL(N,{\bf {C}})/SU(N)$.

\vskip .1in
\noindent{\bf 3. Metric and volume element}

Let us now consider the metrics or distance functions on the relevant spaces. From comparing the action
$S=\int {1\over 4}F^2= {1\over 2} \int \partial_0 A_i^a \partial_0 A_i^a$ 
to a standard form like $\12 g_{\alpha \beta}{\dot Q}^\alpha {\dot Q}^\beta$, 
we see that the relevant metric on $\A$ is given by
$$\eqalign{
ds^2_{\A}~ &=~ \int d^2x~ \delta A^a_i \delta A^a_i ~= 
-8 \int \Tr (\delta A_z \delta A_{\bz} )\cr
&= 8 \int \Tr \left[ D(\delta M M^{-1}) \bD (M^{\dag -1}\delta M^\dag )
\right] \cr}\eqno(7)
$$
where in the last line we have used the parametrization (4) and $D,\bD$ are in 
the adjoint representation.
This is a simple Euclidean metric for $\A$ and the volume element $d\mu (\A )$ 
for this space is the standard Euclidean one, $d\mu (\A )= 
[dA d\bA ] =\prod_x dA(x) d\bA (x)$. 

We now turn to the matrices $M,~M^\dag$; these are
elements of $SL(N,{\bf {C}})$ and 
we have
the Cartan-Killing metric for $SL(N,{\bf {C}})$, viz., $ds^2~=8 {\Tr}(\delta M
M^{-1}~M^{\dagger -1} \delta M^\dagger )$. For $SL(N,{\bf {C}})$-valued fields, we
thus have
$$
ds^2_{SL(N,\bf {C})}  = 8 \int \Tr [(\delta M M^{-1}) (M^{\dagger -1} \delta M^{\dagger})] 
\eqno(8)
$$
We denote the corresponding volume element, the Haar measure, by 
$d\mu (M, M^\dagger )$; we do not need an explicit expression for this at 
this stage. However, from Eq.(7) we can see that $d\mu (\A ) =\det (D \bD )
d\mu (M,M^\dagger )$.

Finally we need to consider $SL(N,{\bf {C}})/SU(N)$. Since this is a coset space
we can start from the $SL(N,{\bf {C}})$-metric and obtain a metric for the 
quotient in a standard way. A simple way to do this is to write an 
$SU(N)$-invariant
version of Eq.(8) by introducing an auxiliary `gauge field' $\alpha$, viz.,
$$
ds^2~= 8 \int \Tr \left[ (\delta M +\alpha M)M^{-1} M^{\dag -1}(\delta M^\dag
+M^\dag \alpha )\right] \eqno(9)
$$
Eliminating $\alpha$ by its equation of motion we get
$$
ds^2_H ~= 2 \int ~\Tr(H^{-1}~\delta H )^2 
~= \int r_{ak} r_{bk} ~\delta \vf^a \delta \vf^b \eqno(10)
$$
where we parametrize $H$ in terms of the real field $\vf^a(x),~
H^{-1}\delta H~= \delta \vf^a ~r_{ak}(\vf ) t_k $.
The corresponding volume element or Haar measure is given  by
$$
d\mu (H) ~= (\det r) [\delta \vf ] \eqno(11)
$$

We are now ready to consider the volume element $d\mu (\C )$ on the configuration
space. This is obtained from $d\mu (\A )$ by factoring out the volume of
gauge transformations. We thus have
$$\eqalign{
d\mu ({\C}) &= {d\mu ({\A})\over vol({\G}_*)}
 = {[dA_z dA_{\bar{z}}]\over vol({\G}_*)} \cr
&= (\det D_z D_{\bar{z}}) {d\mu 
(M, M^{\dagger})\over vol({\G}_*)}
~= (\det D \bD ) d\mu (H)\cr}\eqno(12) 
$$
The problem is thus reduced to calculating the determinant of the two-dimensional
operator $D\bD~^{6,7}$. We have
$$
(\det D \bD) ~= {\rm {constant}}\times\exp \left[ 2c_A ~\S (H) \right]\eqno(13)
$$
where $c_A \delta^{ab} = f^{amn}f^{bmn}$ and $\S (H)$ is the 
Wess-Zumino-Witten (WZW) action for the hermitian matrix field $H$ given by
$$
{\S} (H) = {1 \over {2 \pi}} \int \Tr (\partial H \bar{\partial} H^{-1})
+{i \over {12 \pi}} \int \epsilon ^{\mu \nu \alpha} \Tr ( H^{-1} \partial _{\mu}
H H^{-1} \partial _{\nu}H H^{-1} \partial _{\alpha}H) \eqno(14) 
$$

Eventhough the result (13) is well-known, we shall review it briefly here since
we shall need one of the steps in its evaluation for later purposes. Defining
$\Gamma = \log~\det D\bD$, we have
$$
{\delta \Gamma \over \delta \bA^a}~= -i~\Tr\left[ \bD^{-1}(x,y) 
T^a\right]_{y\rightarrow x} \eqno(15)
$$
Here $(T^a)_{mn}=-if^a_{mn}$ are the generators 
of the Lie algebra in the adjoint representation.
The coincident-point limit of $\bD^{-1}(x,y)$ is, of course, singular and needs
regularization. Since $ d\mu (\C )$ must be gauge-invariant, a gauge-invariant
regularization is appropriate here. With a gauge-invariant regulator such as
covariant point-splitting we have
$$
\Tr \left[ \bD^{-1}_{reg}(x,y)\right]_{y\rightarrow x}~= {2c_A \over \pi}
\Tr \left[ (A -M^{\dag -1} \partial M^\dag )t^a\right] \eqno(16)
$$
Using this result in Eq.(15), and with a similar result for the variation of 
$\Gamma$ with respect to $A^a$, and integrating we get $\Gamma = 2c_A \S (H)$.

Using the expression (13) in Eq.(12), we get the volume element on the
configuration space $\C$ as $^{7,8}$
$$
d\mu (\C )~= d\mu (H)  ~e^{2c_A \S (H)}= [\delta \vf ] (\det r) e^{2c_A \S (H)}
\eqno(17)
$$
The inner product for physical states is given by
$$
\la 1 \vert 2\ra = \int  d\mu (H)  e^{2c_A \S (H)}~\Psi_1^* (H) \Psi_2 (H)
\eqno(18)
$$
Here we begin to see how conformal field theory can be useful; this formula
shows that all matrix elements in $(2+1)$-dimensional $SU(N)$-gauge theory 
can be
evaluated as  correlators of the $SL(N,{\bf {C}})/SU(N)$- WZW model for
the hermitian matrix $H$. (For a general gauge group $G$, we will have a
$G^{\bf {C}}/G$-WZW model, $G^{\bf {C}}$ being the complexification of $G$.)

\vskip .1in
\noindent{\bf 4. Spectrum of $T$, a first look}

Some interesting observations follow from Eq.(17). The integral of $d\mu (\C )$
is the partition function for the WZW-model. We can evaluate this as
$$
\int d\mu (\C ) ~= \int d\mu (H) e^{2c_A \S} ~= \left[ {\det ' \partial \bdel 
\over {\int d^2x}}\right]^{-dim G}   \eqno(19)
$$
where $dimG =N^2-1$ for $G=SU(N)$. Regularizing with a cutoff on the number of
modes we  see that the result is finite; i.e., the
total volume of $\C$ with an appropriate regulator is finite. This is to be 
contrasted with the Abelian case which has $c_A=0$ and where the integral 
diverges for each mode. The ``finiteness" of the volume is clearly a step in 
the right direction as regards the mass gap although we are far from 
any statement of compactness for $\C$.

We can also see, in an intuitive way, how the exponential factor $\exp(2c_A \S )$
can influence the spectrum.
Writing $\Delta E, ~\Delta B$ for the root mean square fluctuations of the 
electric field $E$
and the magnetic field
$B$, we have, from the canonical commutation rules 
$[E_i^a, A_j^b]= -i\delta_{ij}\delta^{ab}$, $\Delta E~\Delta B\sim k$, 
where $k$ is the
momentum variable. This gives an estimate for the energy
$$
{\E}={1\over 2} \left( {e^2 k^2\over\Delta B^2 } +{\Delta B^2 \over e^2}
 \right) \eqno(20)
$$
For low lying states, we minimize ${\E}$ with respect to $\Delta B^2$, 
$\Delta B^2_{min}\sim
e^2 k$, giving ${\E}\sim k$. This is, of course, the standard photon or perturbative
gluon. For the non-Abelian theory, this is inadequate since $\la \H \ra$
involves
the factor $e^{2c_A \S }$. In fact,
$$
\la \H \ra ~= \int d\mu (H) e^{2c_A \S }~ \12 (e^2E^2 +B^2/e^2 ) \eqno(21)
$$
Since $ \S (H) \approx
[ -(c_A /\pi ) \12 \int B (1/k^2 )B +...]$, we see that $B$ follows a Gaussian 
distribution of width $\Delta B^2 \approx \pi k^2 /c_A$, for 
small values of $k$. This Gaussian dominates near small $k$
giving $\Delta B^2 \sim k^2 (\pi /c_A )$. 
In other words, eventhough ${\E}$ is minimized around $\Delta B^2 \sim k$,
probability is 
concentrated around $\Delta B^2 \sim k^2 (\pi /c_A )$. For the expectation
value of the energy,
we then find
${\E}\sim (e^2c_A/2\pi ) +{\O}(k^2)$. Thus the kinetic term in 
combination with 
the measure factor $e^{2c_A \S}$ could lead to a mass gap of order $e^2c_A$. 
The argument is not rigorous; many
terms, such as the non-Abelian contributions to the commutators and $\S (H)$,
have been neglected.
Nevertheless, we expect this to capture the essence of how a mass gap 
could arise.

Before taking up the construction of the Laplacian on $\C$ we need some more
properties of the hermitian WZW-model $^{7,8}$. These can be obtained by comparison 
with the $SU(N)$-model defined by $\exp(k \S (U)),~ U(x)\in SU(N)$. The quantity
which corresponds to $e^{k\S (U)}$ for the hermitian model is $\exp [(k+2c_A)
\S (H)]$. The hermitian analogue of the renormalized level $\kappa = (k+c_A)$
of the $SU(N)$-model is $-(k+c_A)$. Correlators can be calculated from
the Knizhnik-Zamolodchikov equation. Since the latter involves only the 
renormalized level $\kappa$, we see that the correlators of the 
hermitian model (of level $(k+2c_A)$ ) can be obtained from the correlators of 
the (level $k$ ) $SU(N)$-model by the analytic continuation 
$\kappa \rightarrow -\kappa$. For the $SU(N)_k$-model there are the 
so-called integrable representations whose highest weights are limited 
by $k$ (spin $\leq k/2$ for $SU(2)$, for example). Correlators involving 
the nonintegrable representations vanish. For the hermitian model the 
corresponding statement is that the correlators involving nonintegrable 
representations are infinite. (This is not a regularization problem, 
the correlators have singularities at certain values of $k$, the coupling 
constant.) In our case, $k=0$, and we have only one integrable representation
corresponding to the identity operator (and its current algebra descendents).
All matrix elements of the $(2+1)$-dimensional gauge theory being correlators
of the hermitian WZW-model, we must conclude that all wave functions 
(with finite inner product and norm) can be taken to be functions of 
the current
$$
J^a(x) = {c_A \over \pi } \left( \partial H~H^{-1} \right)^a (x)
~= {c_A \over \pi} \left[ iM^{\dag ab} (x) A^b(x) + 
(\partial M^\dag ~M^{\dag -1})^a (x)
\right] \eqno(22)
$$
where $M^{\dag ab}= 2 \Tr (t^a M^\dag t^b M^{\dag -1} )$ is the adjoint 
representation of $M^\dag$.

Although we have gone through the line of reasoning which follows from conformal
field theory, this conclusion is, in the end, not surprising. The Wilson loop 
operator can be written in terms of the current as
$$
W(C)~=~ \Tr ~P ~e^{\oint_C Adz+\bA d{\bar z}}~=
 \Tr ~P~e^{-(\pi /c_A)\oint_C J }\eqno(23)
$$
In principle, all gauge-invariant functions of $(A, \bA )$ can be constructed 
from $W(C)$ and hence it suffices to consider wave functions as functions 
of $J^a$.

We can now start looking at the spectrum of $T= (e^2/2)\int E^2$. Since $T$ is
positive and $E=-i{\delta /\delta A}$, the ground state is given by $\Psi_0 =1$.
This seems too trivial an observation but the key point is that $\Psi_0$ is 
normalizable with the inner product (18).

Consider now an excited state with wave function 
$J^a(x)$.
We have 
$$\eqalignno{
T~J^a(x) &= -{e^2\over 2}\int d^2y {\delta^2 J^a(x)\over 
\delta \bA^b(y) \delta A^b(y)}
~={e^2c_A\over 2\pi} M^{\dag am} \Tr \left[ T^m \bD ^{-1}(y,x) 
\right]_{y \rightarrow x}&(24a)\cr
&= m ~J^a(x)&(24b)\cr}
$$
In Eq.(24a) we encounter the same coincident-point limit as in the calculation
of $(\det D\bD)$. We have used the same regulator and the result (16) to obtain
Eq.(24b), with $m= (e^2c_A/2\pi)$. 

Eq.(24) shows that a state with wave function proportional to $J^a$, say
$\Psi_1 =  \int  d^2x~  f(x)  J^a(x)$ has a mass gap $m$, a rather nice result.
However, $J^a$ is not an acceptable state. The reason is that there are many 
choices for $M$ for a given potential $A$. In particular $M$ and 
$M{\bar V}({\bar x})$, where ${\bar V}({\bar x})$ is an antiholomorphic 
function, lead to the same $A=-\partial M~M^{-1}$. Of course, there are no 
globally defined antiholomorphic functions except the constant and if we 
impose $M\rightarrow 1$ at spatial infinity, this ambiguity can be eliminated. 
However $M$'s corresponding to some $A$'s will have singularities. One can 
eliminate singularities in $M$ by defining it separately on coordinate 
patches and using the antiholomorphic transformation as transition functions,
 i.e., $M_1= M_2 {\bar V}_{12},$ etc., or $H_1 = V_{12} H_2 {\bar V}_{12}$ 
in terms of $H=M^\dag M$. Since this is an ambiguity of choice of field 
variables, the wave functions must be invariant under this. (The ambiguity in 
the choice of $M$ or $H$ and the need for (anti)holomorphic transition 
functions are related to the geometry of $\A$ as a ${\G}_*$-bundle 
over $\C$ and the Gribov problem. For a discussion of these issues, 
see reference 1.)
$J^a$ (or $\Psi_1$) by itself does not satisfy the required condition of 
``holomorphic invariance"; we need at least two $J$'s. 
We should thus consider the action of the kinetic energy operator $T$ on 
two $J$'s or more generally we need $T$
as an operator on any function of $J$'s.

\vskip .1in
\noindent{\bf 5. The Laplacian and $T$}

In constructing the operator $T$, first let us consider the change of variables 
from $A,\bA$ to $M,M^\dag$. The metric on $\A$ can be written as
$$
ds^2_{\A}~= -8 \int \Tr (\delta A \delta \bA ) = \int g_{a{\bar b}}(x,y) 
\delta\theta^a(x) \delta {\bar\theta}^b(y)~+~ h.c.\eqno(25)
$$
where $\theta,~{\bar \theta}$ are parameters defining $M$ and $M^\dag$
respectively and
$$\eqalignno{
g_{a{\bar b}}(x,y) ~&= 2 \int_{u,v} \partial_u [\delta (u-x)R_{ar}(u)]
M_{kr}(u) M^\dag_{sk}(v) {\bar \partial}_v [\delta (v-y) R^*_{bs}(v)]&(26)\cr
M^{-1}\delta M~ &= \delta \theta^a R_{ab}(\theta ) t_b, ~~~~~ 
\delta M^\dag M^{\dag -1} ~= \delta {\bar \theta}^a R^*_{ab}({\bar \theta})t_b
&(27)\cr}
$$

The metric on $\A$ is a K\"ahler metric since $ds^2 = \delta_A \delta_{\bA}[-8\int
\Tr(A\bA )]$. The Laplacian $\D$ has the general form
$$
\D~= g^{-1}\left( \partial_{\bar a} g^{{\bar a}a}g \partial_a + 
\partial_a g^{a{\bar a}}g \partial_{\bar a}\right) \eqno(28)
$$
where $g=\det(g_{a{\bar a}})$.
Eq.(26) then leads to
$$\eqalign{
T~= -{e^2\over 2} \D = {e^2\over 4} \int_x e^{-2c_A \S (H)} \bigl[ &{\bar G}\bp_a(x)
K_{ab}(x) e^{2c_A\S (H)} Gp_b(x) ~+\cr
& Gp_a(x) K_{ba}(x) e^{2c_A\S (H)} {\bar G}\bp_b(x)\bigr]\cr}\eqno(29)
$$
where $p_a$ generates right-translations on $M$ and $\bp_a$ generates 
left-translations on $M^\dag$, i.e.,
$$\eqalign{
[p_a(x), M(y)] ~&=M(y) (-it_a)~\delta (y-x)\cr
[\bp_a (x), M^\dag (y) ] ~&= (-it_a) M^\dag (y) ~\delta (y-x)\cr}\eqno(30)
$$
Further, $K_{ab}= 2 \Tr (t_a H t_b H^{-1})$ is the adjoint representation of 
$H$ and $Gp_a(x)=\int_u G(x,u) p_a(u)$ (and similarly for $\bG \bp_a$).

Rather than constructing the Laplacian as above, we can also write $T\Psi =
(-e^2/2) ({\delta^2 \Psi }/{\delta A \delta \bA})$ and make the change of 
variables to $M,M^\dag$. This gives the expression
$$
T~= {e^2\over 2}\int_x K_{ab}(x) \bG \bp_a(x) Gp_b(x) \eqno(31)
$$
For a finite dimensional K\"ahler manifold, we have the identity 
$\partial_{\bar a}(g^{{\bar a}a}g)=0$ and this suffices to prove that the 
two forms of $T$, viz., Eqs.(29,31), are in fact identical. In our case, 
eventhough the metric is K\"ahler, the equality of the two expressions 
does not immediately follow since the space is infinite dimensional; we 
need to regularize these expressions. A regularized expression preserving 
gauge and holomorphic invariance is given by $^2$ 
$$
T~= {e^2\over 2} \int_{u,v}\Delta_{ab}(u,v) \bp_a(u) p_b(v)\eqno(32)
$$
$$\eqalign{
\Delta_{ab}(u,v)~&= \int_x {\bar {\G}}_{ma}(x,u) K_{mn}(x) {\G}_{nb}
(x,v)\cr
{\bar {\G}}_{ma}(x,u) ~&= {\bar G} (x,u) \bigl[\delta_{ma}~-~
e^{-\vert x-u\vert^2/\epsilon}~
\bigl( K(x,{\bar u}) K^{-1}(u,{\bar u})\bigr)_{ma}\bigr]\cr
{\G}_{nb}(x,v)~&= G(x,v) \bigl[ \delta_{nb}~-~
e^{-\vert x-v\vert^2/\epsilon}~
\bigl( K^{-1}(v,{\bar x}) K(v,{\bar v}) \bigr)_{nb} \bigr] \cr}\eqno(33)
$$
The regularized Green's functions ${\G}, ~{\bar{\G}} \rightarrow G, 
~\bG$ (times the Kronecker delta )
as the regularizing parameter $\epsilon \rightarrow 0$ and ${\G}(x,x)=
{\bar{\G}}(x,x) =0$ for finite $\epsilon$. 

One can check that, for $T$ defined 
by Eqs.(32,33), we have
$$\eqalign{
T~={e^2\over 4} \int_x e^{-2c_A \S (\epsilon ,H)} \bigl[ &({\bar{\G}}\bp)_a(x)
K_{ab}(x) e^{2c_A\S (\epsilon ,H)} ({\G}p)_b(x) ~+\cr
& ({\G}p)_a(x) K_{ba}(x) e^{2c_A\S (\epsilon ,H)} ({\bar{\G}}\bp)_b(x)
\bigr]\cr}\eqno(34)
$$
where $\S (\epsilon,H) =\S (H) +{\O}(\epsilon )$. 

The above considerations apply to the space $\A$. The restriction to $\C$ is,
however, straightforward. Using $M=U\rho$, the operator $p_a$ can be written
as
$$
p_a~= -i R_{ab}^{-1}{\delta \over \delta \theta^b}~= (1+M)_{ab}^{-1} 
(\alpha_b +I_b)\eqno(35)
$$
where $\alpha_b$ generates right-translations on $\rho$ and $I_b$ generates 
translations on $U$. (There is a similar expression for $\bp_a$.) On 
gauge-invariant functions which are independent of $U$, $I_b=0$. Setting
$I_b=0$ in Eqs.(32,34), we get the required expressions on $\C$. When
$I_b=0$, $p_a=-ir_{ab}^{-1} {\delta \over \delta \vf^b}$ and $ \bp_a =K_{ab}p_b$.
With this understanding and using Eq.(32) for $T$ we 
find
$$
T~\Psi (J)= m \left[ \int_w \omega^a(w){\delta \over \delta J^a(w)}~+\int_{w,z}
\Omega^{ab}(w,z) {\delta \over \delta J^a(w) }{\delta \over \delta J^b(z)}\right]
\Psi(J) \eqno(36)
$$
$$\eqalignno{
\omega^a(w)&= -if^{abc} K_{cs}(w)\left[\partial_w \Delta_{bs}(y,w)
\right]_{y\rightarrow w} &{}\cr
&=J^a(w) ~+{\O}(\epsilon )&(37a)\cr
\Omega^{ab}(w,z)&= -K_{an}(w)\left[ \left({c_A\over \pi}\partial_z \delta^{bm} 
+if^{bmk}J^k(z)\right) \partial_w\Delta_{mn}(z,w)\right]&{}\cr
&= \left[ {\delta^{ab}\over {\pi (w-z)^2}}- {if^{abc}J_c(z)\over {\pi (w-z)}}
\right]~+
{\O}(\epsilon )&(37b)\cr}
$$
We see that as $\epsilon \rightarrow 0$, the first term in $T$ gives the 
number of $J$'s in $\Psi (J)$ while the second replaces pairs of $J$'s by 
the lowest terms of the
operator product expansion for currents in the 
WZW-model.

We now return to the consideration of the states. The quantity 
$\bdel J^a (x) \bdel J^a(x)$ has both gauge and holomorphic invariance and
we can construct a physical state
$$
\Psi_2 (J) = \int_x f(x) \left[ \bdel J_a(x) \bdel J_a(x) + {c_A{\rm {dim}}G \over \pi^2} 
\partial_x
\bdel _x \delta (x-y) \vert_{y\rightarrow x} \right] \eqno(38)
$$
The second (c-number) term in this expression is necessary to orthogonalize 
this with respect to the ground state. It is also what is needed for normally
ordering the term $\bdel J^a (x) \bdel J^a(x)$.
From the above formula for the action of $T$ we find
$$
T~\Psi_2 = 2m~\Psi_2 \eqno(39)
$$
This is the lowest excited eigenstate of $T$.
One can form more general combinations, for example,
$$\eqalign{
\Psi &= \int f(x_1,x_2) :\Tr [\bdel J(x_1) U(x_1,x_2) \bdel J(x_2)]: \cr
& +\int f(x_1,x_2,x_3) :\Tr[ \bdel J(x_1) U(x_1,x_2) \bdel J(x_2) U(x_2,x_3) 
\bdel J(x_3) U(x_3,x_1)]: +\cdots\cr}\eqno(40)
$$
where $U(x,y)= K(x, {\bar y})K^{-1}(y, {\bar y})$. By requiring that this be 
an eigenstate of $T$, we obtain a hierarchy of coupled equations for the 
functions $f(x_1,x_2),~f(x_1,x_2,x_3)$, etc.. The solutions will give 
series of eigenstates and eigenvalues, a series for $f(x_1,x_2)$, a series 
for $f(x_1,x_2,x_3)$, etc.; we expect these to be analogous to Regge 
trajectories.

So far we have talked about the spectrum of the kinetic energy operator $T$. 
How do we include the potential term ? 
We expect that this can be done perturbatively, $1/e^2$ being the 
expansion parameter. 
Since $m=(e^2c_A/2\pi )$, this will be an expansion in $1/m$, say in 
powers of $p/m$ where $p$ is a typical momentum. For example, upto the 
first order in perturbation theory we find
$$
{\H}J^a =(T+V)J^a \approx \left( m~+ {p^2\over 2m}\right) J^a  ~+
{\O}\left({1\over m^2}\right)\eqno(41)
$$
We get the first correction to the energy beyond the
mass $m$ in a small momentum expansion. Since the theory is Lorentz invariant
higher order corrections are expected to sum upto the relativistic expression
$(p^2+m^2)^{1\over 2}$. The situation is similar to what occurs for solitons;
there again one finds a similar expansion which sums upto the relativistic
formula $^9$.)  
Notice that the state $\Psi_2$, for example, is degenerate, having the same 
eigenvalue for $T$ for all $f(x)$ ( and similarly for the more
general states of Eq.(40)). This degeneracy is lifted by the potential term.
The inclusion of the potential term is thus necessary to obtain a proper
hierarchy of equations for the functions $f(x_1,x_2),~f(x_1,x_2,x_3)$, etc..
\vskip .1in
\noindent{\bf 6. Concluding remarks}

From what we have said so far it should be clear that this is a fruitful 
line of investigation and there are many remaining questions of interest. 
The construction of a complete set of eigenstates for $T$, the inclusion 
of the potential term and the derivation of the hierarchy of equations 
for $f(x_1,x_2),~f(x_1,x_2,x_3)$, etc. are of prime importance. We are 
currently investigating these questions. Another very interesting 
possibility is the following. Once we have the Hamiltonian $\H$ as 
an operator on functions on $\C$, we can obtain a functional integral, 
starting from $e^{-i \H t}$, which is defined directly on $\C$ without 
the need for gauge fixing. There are also interesting generalizations 
of the pure gauge theory, such as the inclusion of quarks or a 
Chern-Simons mass term, which can be investigated using our approach.

\vskip .1in
\noindent{\bf Acknowledgements}

This work was supported in part by the National Science Foundation
grant PHY-9322591 and by the Department of Energy grants 
DE-FG02-90ER40542 and DE-FG02-91ER40651-Task B.
\vskip .1in
\noindent{\bf References}
\vskip .1in
\item{1.} D. Karabali and V.P. Nair, {\it Nucl.Phys.} {\bf B464} (1996) 135; {\it Phys.
Lett.} {\bf B379} (1996) 141.
\item{2.} D. Karabali, Chanju Kim and V.P. Nair, Preprint CCNY-HEP 96/12. 
\item{3.}  A.M. Polyakov, {\it Nucl.Phys.} {\bf B120 } (1977) 429.
\item{4.} R.P. Feynman, {\it Nucl.Phys.} {\bf B188} (1981) 479. 
\item{5.} I.M. Singer, {\it Phys.Scripta} {\bf T24} (1981) 817; 
{\it Commun. Math. Phys.} {\bf 60} (1978) 7; M. Atiyah, N. Hitchin and 
I.M. Singer, {\it Proc. Roy. Soc. Lond.} {\bf A362} (1978) 425; P.K. Mitter and
C.M. Viallet, {\it Phys. Lett.} {\bf B85} (1979) 246;
{\it Commun. Math. Phys.} {\bf 79} (1981) 457; M. Asorey and P.K. Mitter, 
{\it Commun. Math. Phys.} {\bf 80} (1981) 43; O. Babelon and C.M. Viallet, 
{\it Commun. Math. Phys.} {\bf 81} (1981) 515; {\it Phys. Lett.} {\bf B103} (1981) 45; P. Orland, NBI-CUNY preprint, hep-th 9607134.
\item{6.}  A.M. Polyakov and P.B. Wiegmann, {\it Phys.Lett.} {\bf B141} 
(1984) 223; D. Gonzales and A.N. Redlich, {\it Ann.Phys.(N.Y.)} 
{\bf 169} (1986) 104;
B.M. Zupnik, {\it Phys.Lett.} {\bf B183} (1987) 175; 
G.V. Dunne, R. Jackiw
and C.A. Trugenberger, {\it Ann.Phys.(N.Y.)} {\bf 149} (1989) 197;
D. Karabali, Q-H Park, H.J. Schnitzer and Z. Yang, {\it Phys.Lett.}
{\bf B216} (1989) 307; D. Karabali and H.J. Schnitzer, {\it Nucl.Phys.}
{\bf B329} (1990) 649.
\item{7.} K. Gawedzki and A. Kupiainen, {\it Phys.Lett.} {\bf B215} (1988) 119;
{\it Nucl.Phys.} {\bf B320} (1989) 649.
\item{8.} M. Bos and V.P. Nair, {\it Int.J.Mod.Phys.} {\bf A5} (1990) 959.
\item{9.} J. Goldstone and R. Jackiw, {\it Phys. Rev.} {\bf D11} (1975) 1486;
R. Jackiw, {\it Rev. Mod. Phys.} {\bf 49} (1977) 681;
J.L. Gervais and B. Sakita, {\it Phys. Rev.} {\bf D11} (1975) 2943;
J.L. Gervais, A. Jevicki and B. Sakita, {\it Phys. Rev.} {\bf D12} (1975) 1038;
N. Christ and T.D. Lee, {\it Phys. Rev.} {\bf D12} (1975) 1606.

\end